\documentclass[11pt,tightenlines,onecolumn, aps, prd, nofootinbib, superscriptaddress, showkeys, showpacs]{revtex4-1}

\usepackage{latexsym}
\usepackage{amssymb}
\usepackage[dvips]{graphicx}
\usepackage{color}
\usepackage{graphicx}

\usepackage{amsmath, amsfonts, amssymb, mathrsfs}
\usepackage{amsthm}
\usepackage{tensor}
\usepackage{multirow}
\usepackage{bbm}
\usepackage[geometry]{ifsym}

\usepackage{tensor}
\begin{document}

\title{Galaxy rotation curves in the light of the Spinor Theory of Gravity}

\author{M. Novello}
\email{novello@cbpf.br}
\affiliation{Centro de Estudos Avan\c{c}ados de Cosmologia (CEAC/CBPF) \\
 Rua Dr. Xavier Sigaud, 150, CEP 22290-180, Rio de Janeiro, Brazil.}
\author{A. E. S. Hartmann}
\email{aedshartmann@uninsubria.it}
\affiliation{Dipartamento di Scienza e Alta Tecnologia, Universit{\`a} degli Studi dell’Insubria,\\
Via Valleggio 11, 22100 Como, and INFN Sez di Milano, Italy.}
\author{E. Bittencourt}
\email{bittencourt@unifei.edu.br}
\affiliation{Universidade Federal de Itajub\' a, Av.\ BPS 1303, Itajub\' a-MG, 37500-903, Brazil}
\
\date{\today}

\vspace{0.50cm}

\begin{abstract}
We analyze the recently obtained static and spherically symmetric solution of the Spinor Theory of Gravity (STG) which, in the weak field limit, presents an effective Newtonian potential that contains an extra logarithmic behavior. We apply this solution to the description of the galaxy rotation curves finding an interesting analogy with the dark matter halo profile proposed by Navarro, Frenk and White. 
\end{abstract}

\pacs{95.30.Sf, 04.20.-q, 98.80.-k}

 \maketitle
%\section{Introduction}
Recently \cite{novhart20a,novhart20b}, some of us have analyzed the hypothesis according to which gravity is controlled by spinor fields that combine in an effective metric. In that proposal, the main hypothesis of General Relativity (GR) that states that gravity may be interpreted as nothing but a modification of the geometry of the underlying space-time is followed. However, the further hypothesis of GR that states that the metric satisfies an independent dynamical equation controlling the action of matter on the modifications of the geometry is avoided. In STG, it is assumed that the geometry is an efficient way to describe the gravitational interaction between bodies, but it should not have a dynamical equation by its own; the space-time geometry should be interpreted as an effective metric in the way considered by Ref. \cite{novellovisservolovik}. In other words, we are assuming that the geometry that represents the gravitational interaction is not a fundamental object by itself,
but instead it is the result of the behavior of other existing fields founded in Nature.
In the past, different authors have considered gravity in terms of a spin-2 field in a non-observable auxiliary Minkowski space-time structure (see, for instance, Ref.\ \cite{feynman}). In \cite{novhart20a,novhart20b}, a new program to consider an intimate connection between gravity and Fermi (weak) interaction has begun. As previously discussed in these papers, the main hypothesis of the Spinor Theory of Gravity relies upon a suggested connection between weak interactions and gravity,
namely their high degree of universality, and the dimensional equivalence of their coupling
constants in natural units system. The same hypothesis has been also analyzed in distinct contexts,
as for instance in \cite{onofrio1, onofrio2}.

The general framework of the STG concerns two spinor fields (which we shall call gravitational-neutrinos or g-neutrinos, for short). Here, we shall limit our analysis to the case in which one of these fields is absent, but the remaining spinor field possesses self-interaction provided by the Heisenberg potential \cite{heisenberg,heisenberg2,nambu}. Thus, the Dirac-Heisenberg equation satisfied by the g-neutrino takes the form
\begin{equation}
 i\gamma^{\mu}
\nabla_{\mu} \, \Psi \,-\, 2 s \, (A + iB \gamma_{5}) \, \Psi = 0,
\label{domingo2}
\end{equation}
with $ A \equiv \overline{\Psi} \, \Psi$ and  $ B \equiv i \, \overline{\Psi} \gamma_{5} \, \Psi$. The coupling constant $s$ has, in the natural units system $c=\hbar=1$, dimesion of lenght squared. The associated gravitational effective metric \cite{novhart20a} exhibits an intimate connection between weak interaction and gravity and it is given by
\begin{equation}
g_{\mu\nu} =  \eta_{\mu\nu} - \kappa \, \Delta_{\mu} \, \Delta_{\nu},
\label{7maio21}
\end{equation}
where $\kappa$ is the Einstein's gravitational constant. The point of contact between Fermi's interaction and gravity comes through the weak current $ \Delta_{\mu}$ defined as
 \begin{equation}
\Delta_{\mu} =  ( J_{\mu} - I_{\mu} ) \left(\frac{g_{w}}{J^{2}}\right)^{1/4}\,.
\label{18abril4}
\end{equation}
The vector and axial currents of $ \Psi$ are, respectively, represented by $J_{\mu}$ and $I_{\mu}$. We also denote $J^2 = J_{\mu} \, J^{\mu}$ and $g_w$ is the weak constant. In natural units, both $g_w$ and $\kappa$ have dimesion of lenght squared.

In the search of a static and spherically symmetric solutions of STG, we set for a self-interacting g-neutrino, (see Ref. \cite{novhart20a}) $\Psi(r) = f(r) \,\Psi^0,$ such that the constant spinor can be decomposed as
\begin{align*}
\Psi^0 &= \left(\begin{array}{l} \phi\\
\eta\\
\end{array}
\right)\,.
\end{align*}
Moreover, we consider the case in which $\sigma_{1} \phi = \phi,$ and $\sigma_{1} \eta = \eta.$ Thus, after a change of coordinates, it follows for the light-like vector $\Delta_{\mu}$ that $ \Delta_{2} = \Delta_{3} = 0$ (see details in Ref.\ \cite{novhart20a}). Finally, we arrive at the effective gravitational metric -- experienced by all kind of matter in the case of a spherically symmetric and time-independent configuration -- expressed in terms of the infinitesimal line element
\begin{equation}
ds^2 = \left[ 1 - \frac{1}{ r a_{0} - 2 r\, \lambda \, \ln \left(\frac{r}{r_{0}}\right)} \right] \,dT^2 - \left[ 1 - \frac{1}{ r a_{0} - 2 r\, \lambda \, \ln \left(\frac{r}{r_{0}}\right)} \right]^{-1} \, dr^2 - r^2 \, d\Omega^2,
\label{2dez1}
	\end{equation}
where $a_0$, $\lambda$ and $r_{0}$ are parameters depending on the self-coupling constant $s$ and on the solution $\Psi^0$ of the g-neutrino \cite{novhart20a}. In the absence of self-interaction ($\lambda\rightarrow0$), this reduces to the Schwarzschild metric of GR through the identification $a_{0} = 1/2m$, where $m$ is the object mass. In the weak field limit, we can read out the effective gravitational potential as
\begin{equation}
\label{eff_phi}
\Phi_{{\rm STG}}(r)= -\frac{Gm}{r\left[ 1 - 4Gm \lambda \, \ln \left(\frac{r}{r_{0}}\right)\right]}.
\end{equation}
Thus, the galaxy rotation curves can be obtained through the virial theorem as
\begin{equation}
\label{v2-phi}
V^2_{{\rm STG}}=r\frac{d\Phi_{{\rm STG}}}{dr}.
\end{equation}

In general, the expression for $V^2_{{\rm STG}}$ is rather complicate and, on top of that, its correct behavior near the origin $r=0$ is still under investigation, since the baryonic mass has independent components like bulge, disk, gas, etc and each of them has a collection of possible profiles according to the literature \cite{gilmore}. However, the behavior of $V^2_{{\rm STG}}$ far from the origin can roughly provide the pattern of the rotation curves if the baryonic mass is constant. Furthermore, we expect that the coupling constant $\lambda$ be small in order to recover the particle physics results when the gravitational effects are negligible. Together, these hypotheses lead Eq.\ (\ref{eff_phi}) to the following approximation
\begin{equation}
\label{eff_phi_app}
\Phi_{{\rm STG}}(r)\approx -\frac{Gm}{r}\left[ 1 + 4Gm\lambda \, \ln \left(\frac{r}{r_{0}}\right)\right].
\end{equation}
The substitution of this expression into Eq.\ (\ref{v2-phi}) yields
\begin{equation}
\label{v2-stg}
V^2_{STG}\approx\frac{Gm}{r}+\frac{4G^2m^2\lambda}{r}\left[-1+\ln\left(\frac{r}{r_0}\right)\right].
\end{equation}

At this point, it is worth to mention that some phenomenological approaches used to describe the circular velocities of stars around the host galaxy assuming a dark matter component also gives a logarithmic leading term at large values of the radial coordinate \cite{bajkova}. One of the most accepted dark matter halo profile showing this behavior is the model due to Navarro, Frenk \& White (NFW) \cite{navarroetal}, for which the rotation velocity becomes
\begin{equation}
\label{v2-nfw}
V^2=\frac{ G M(r)}{r}+\frac{4\pi G r_s^3\rho_{s}}{r}\left[-\frac{r}{r+r_s}+\ln\left(1+\frac{r}{r_s}\right)\right],
\end{equation}
where $M(r)$ is the baryonic mass enclosed within a sphere with radius $r$. The free parameters of this model are the scale radius and the characteristic density, here denoted by $r_s$ and $\rho_s$, respectively.

For large enough radius such that $r\gg r_s$, a straightforward comparison between Eq.\ (\ref{v2-stg}) and the approximate version of Eq.\ (\ref{v2-nfw}) gives the same Newtonian component, besides the identification
\begin{equation}
\label{comp_par}
r_0=r_s\quad \mbox{and}\quad \lambda= \frac{\pi  r_s^3\rho_s}{Gm^2}.
\end{equation}
This means that the phenomenological behavior observationally tested given by Eq.\ (\ref{v2-nfw}) can also be reproduced from a more fundamental theory if one understands the geometrical manifestation of gravity as a consequence of the self-interacting spinor field in the lines of STG.

For the sake of illustration, let us consider the case of Milky Way, which is a relative normal spiral galaxy, whose the baryonic mass is approximately $6\times10^{10}M_{\odot}$. In order to reproduce its rotation curve using the NFW profile, the total mass of the dark matter halo is assumed to be of the order $\sim 10^{12}{\rm M}_{\odot}$ (see Ref.\ \cite{wang} and references therein)\footnote{Although the error bars can be large, we are only interested in the magnitude of the free parameters in this preliminary analysis.}. For convenience, the astrophysicists usually rewrite the
scale radius and the characteristic density in terms of $M_{200}$ (the total mass of the particles within a sphere with virial
radius $r_{200}$ , which in turn is defined such that the density of the dark matter halo is 200 times the critical density of
the universe $\rho_{\rm crit}$ ) and the halo concentration parameter $c \equiv r_{200} /r_s$ through the formulas \cite{mo,alefe}
\begin{equation}
\label{eq_r_rho}
r_s=28,8\left( \frac{{\rm M}_{200}}{10^{12}\,h^{-1}\,{\rm M}_{\odot}}\right)^{0,43}\, {\rm kpc},\quad \mbox{and}\quad
\rho_s=\frac{200}{3}\frac{c^3\rho_{{\rm crit}}}{\ln(1 + c) - \frac{c}{1 + c}},
\end{equation}
where h is the reduced Hubble constant. The concentration parameter can also be given by \cite{dutton,alefe}
\begin{equation}
\label{eq_c}
c=10^{0,905}\left( \frac{{\rm M}_{200}}{10^{12}h^{-1}M_{\odot}}\right)^{-0,101}.
\end{equation}

If we adopt ${\rm M}_{200}$ at the order of the dark matter halo mass, using $\rho_{{\rm crit}}=143,84 M_{\odot}/kpc^3$ and $h=0,671$ \cite{planck} into Eqs.\ (\ref{eq_c}) and (\ref{eq_r_rho}),
we get $r_s\sim24.2\, \mbox{kpc}$. By substitution of this result into Eq.\ (\ref{comp_par}), we determine immediately $r_0$ and also obtain
\begin{equation}
\label{comp_par_milky}
\lambda\sim2.8\times 10^{-8}\,\mbox{eV}, 
\end{equation}
where we used the conversion factor $1\,\mbox{pc}^{-1}\sim2.6\times10^{-11}\,\mbox{eV}$. Of course, it is a first attempt to explain the galaxy rotation curve in the realm of the STG, but this works as a consistency test for our model.
In fact, we see that these values are in agreement with the assumptions made before about the domain of validity of Eq. (8) and, in particular, it does not alter the particle physics once energy scale of self-interaction is very small and, consequently, hard to be detected nowadays. Notwithstanding, the most adequate procedure to find the values for the best fit of $r_0$ and $\lambda$ is taking a large sample of galaxies (or doing numerical simulations, and then) confronting the results with other dark matter profiles and the particle physics, but this is left for further investigation.

We conclude that in the realm of the STG there may be room for the explanation of the observed profile of the galaxy rotation curves with no need of assuming the hypothesis of the existence of a new unknown form of invisible matter in the universe. Of course, the hypothesis of a dark matter component in the context of RG concerns other effects like the gravitational lensing and the growth of large scale structure, which have not been claimed to be solved solely with the discussion developed along this letter and should be matter for future work. However, the possibility of having an alternative way to describe part of such weird component of the universe in terms of a gravitational theory grounded upon particle physics should be taken into account seriously.

\section*{Acknowledgements}
M. Novello thanks a fellowship from CNPq and FAPERJ.

 \end{document}